\documentclass[aps,prd,showpacs,twocolumn]{revtex4}
\usepackage{graphicx}
\usepackage{dcolumn}
\usepackage{bm}

\begin{document}

\def\be{\begin{equation}}
\def\ee{\end{equation}}
\def\bd{\begin{displaymath}}
\def\ed{\end{displaymath}}
\def\ba{\begin{eqnarray}}
\def\ea{\end{eqnarray}}
\def\C{\rm C}
\def\E{\rm E}
\def\J{\rm J}
\def\L{\rm L}
\def\M{\rm M}
\def\P{\rm P}
\def\S{\rm S}

\title{\Large Charmonium production in $p\bar{p}$ annihilation:\\
              Estimating cross sections from decay widths
} 
\author{
A.Lundborg,$^a$\footnote{Email: agnes.lundborg@tsl.uu.se}
T.Barnes$^b$\footnote{Email: tbarnes@utk.edu}
and
U.Wiedner$^a$\footnote{Email: ulrich.wiedner@tsl.uu.se} 
}
\affiliation{
$^a$Uppsala University, Department of Radiation Sciences,
SE-75121 Uppsala, Sweden\\
$^b$Department of Physics and Astronomy, University of Tennessee,
Knoxville, TN 37996, USA\\
Physics Division, Oak Ridge National Laboratory,
Oak Ridge, TN 37831, USA}

\date{\today}

\begin{abstract}
The cross sections for the charmonium production processes
$p\bar p\to m\Psi $, where $m$ is a light meson
and $\Psi$ is a charmonium state, are of great interest for the
planned $p\bar{p}$ experiment PANDA. In this paper we estimate
these cross sections using known results for the decays $\Psi \to mp\bar p$,
which are related to these reactions by crossing.
In lieu of detailed experimental data on the decay Dalitz plots,
we assume a constant amplitude as a first approximation; this implies
a simple relation between the cross sections and decay widths.
The single measured exclusive cross section of this type
is $p\bar p\to\pi^0 J/\psi$, which was reported by E760 to be
$130\pm 25~pb$ near $\sqrt{s}=3.5$-$3.6$~GeV. In comparison,
our constant amplitude estimate is about $300~pb$ at this energy.
This suggests that our approach is useful as a simple estimate, 
but should be refined through detailed modeling of resonances and other
energy-dependent effects in the experimental $\Psi \to mp\bar p$ Dalitz plots.
\end{abstract}


\maketitle

\section{Introduction}
The proton-antiproton annihilation experiment PANDA, which is planned for the
FAIR facility at GSI, is scheduled to begin data taking
in the beginning of the next decade. PANDA is an internal 
fixed target experiment, with a
design luminosity of $2\cdot 10^{32}$ cm$^{-2}$s$^{-1}$ and a momentum range
of p$_{{\bar p}\, lab}=1.5$-$15$ GeV/c. This covers a mass range in which
many interesting hadron resonances are anticipated, including 
glueballs, light hybrids, charmonia and charmonium hybrids 
\cite{PandaTechnicalProgress}.

The physics goals of PANDA include a detailed investigation of the
spectrum of charmonia and charmonium hybrids, including determinations
of masses, widths, quantum numbers and decay properties.
Since the known charmonium spectrum is largely characterized by 
clear, well separated states, unlike the broad, overlapping resonances 
of light quark hadron spectroscopy, it is anticipated that this will 
be a relatively clean sector in which to search for higher-mass excitations 
such as hybrids and other unusual hadron resonances.

Studies of conventional meson spectroscopy using proton-antiproton 
collisions have a considerable advantage over electron-positron colliders,
in that {\it all} nonexotic quantum numbers can be accessed in formation 
reactions. 
Although ${\J}^{\P\C}$-exotics cannot be made in formation
in $p\bar p$ annihilation, they can be made in associated production 
processes such as $p\bar p \to \pi H_c$ (where $H_c$ is a 
${\J}^{\P\C}$-exotic charmonium hybrid).

\vfill\eject

Production of the very characteristic ${\J}^{\P\C}$-exotic charmonium hybrids
is one of the main goals of the PANDA project. These states are predicted
to lie at rather high masses; the ${\J}^{\P\C} = 1^{-+}$ state, which is 
expected to be the lightest exotic charmonium hybrid, is predicted 
by the flux tube model
\cite{Isgur:1984bm,Merlin:1985mu,Merlin:1986tz,Barnes:1995hc}
and lattice QCD 
\cite{Liao:2002rj,Juge:2003qd,Bali:2003tp}
to have a mass in the range 4.2-4.4 GeV.
This exotic resonance could be produced at PANDA together with a light meson
$m$, such as an $\eta$ or $\pi^0$. (Both isospins are allowed in
$p \bar p \to m\Psi $ because $p\bar p$ is a superposition of I=0 and I=1.)

Numerical estimates of charmonium and charmonium hybrid
production cross sections are evidently crucial for the PANDA physics program,
both to formulate detection strategies and to evaluate luminosity requirements, 
as well as for detailed detector simulations with theoretically preferred 
final states. Unfortunately there are few estimates of these 
cross sections at present, and only one has been measured, 
$p\bar{p}\to\pi^0 J/\psi $. (This result was reported by the E760 
collaboration \cite{Armstrong:1992ae}.) 
The purpose of this paper is to derive and evaluate simple estimates 
of these cross sections from measurements of charmonium partial widths
to $m p\bar p$ three-body final states.

The approach we follow in this paper is to use a measured charmonium 
partial width into a three-body  $m p \bar p$  final state,
$\Gamma_{\Psi\to m p\bar p}$,
to estimate the corresponding production
cross section $\sigma_{p\bar p \to m\Psi}$.
The (presumably complicated) quark-gluon dynamics
is subsumed in the unknown amplitude $\cal{M}$, which is
common to both processes.

\vfill\eject

\section{Relating widths to cross sections}

\subsection{Three-body decay $\Psi\to m p\bar p$}

\begin{figure}[ht]
\begin{center}
\includegraphics[width=0.4\linewidth]{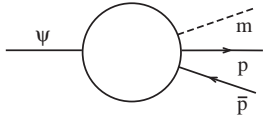}
\caption{The three-body charmonium decay process
$\Psi \to m p\bar p $.}
\label{fig:3body}
\end{center}
\end{figure}

The three-body partial width of an initial charmonium state $\Psi$
is an average over initial and sum over final polarizations of a squared decay
amplitude, multiplied by three-body phase space.
The differential decay rate, in the form familiar as a Dalitz plot, 
is
\begin{equation}
d\Gamma_{\Psi\to m p\bar p}
 =\frac{1}{2S_{\Psi}+1}
\frac{1}{(2\pi)^3}\frac{1}{32 M_{\Psi}^3} 
\Big\{ 
\sum |\mathcal{M} |^2
\Big\}
\, dm^2_{m p}\, dm^2_{p\bar p} \ .
\label{eq:width}
\end{equation}
Here $S_X$ is the spin of particle $X$,
$m^2_{ij} = (p_i + p_j)^2$ is the invariant mass squared 
of the two-particle system
$(i,j)$ \cite{PDG}, and the squared amplitude in this formula 
is summed over all $4(2S_{\Psi}+1)(2S_{m}+1)$ initial and final 
polarization states. (This summation convention is convenient for 
relating decays to cross sections.) 

As a simple approximation one can assume a constant decay amplitude
$\cal{M}$, in which case 
the decay width is proportional to the area $A_D$ of the Dalitz 
plot;
\begin{equation}
\Gamma_{\Psi\to m p\bar p}
 =\frac{1}{2S_{\Psi}+1}
\frac{1}{(2\pi)^3}\frac{1}{32 M_{\Psi}^3} 
\Big\{
\sum |\mathcal{M} |^2
\Big\}
\, A_D 
\label{eq:width_constAmp}
\end{equation}
where
\begin{equation}
A_D = 
\int \!\!\! \int
dm^2_{m p}\, dm^2_{p\bar p} \ .
\label{eq:A_D_defn}
\end{equation}

\subsection{Two-body production reaction $p\bar p\to m\Psi$}

\begin{figure}[ht]
\begin{center}
\includegraphics[width=0.4\linewidth]{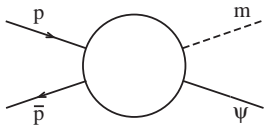}
\caption{The related associated charmonium production process 
$p\bar p \to m \Psi$.}
\label{fig:2body}
\end{center}
\end{figure}

In the unpolarized two-body production cross section 
the squared amplitude is again 
averaged over initial and summed over final states. 
The unpolarized differential cross section is given by \cite{PDG}
\begin{equation} 
\frac{d\sigma}{dt}\bigg|_{p\bar p \to m\Psi} =
\frac{1}{256\pi}\frac{1}{|p_{p\, cm}|^2} s^{-1}
\Big\{
\sum |\mathcal{M}|^2
\Big\} 
\end{equation}
where $t$ is the Mandelstam variable $t = (p_p-p_m)^2$. 
A factor of 1/4 has been included for the average over the four initial
$p\bar p$ polarization states.

The total cross section involves an integral over $t$ between the limits
\be
t_{\pm}
=
m_p^2 + m_m^2 - 2 (E_pE_m \mp p_p p_m)_{cm}.
\ee
Assuming a constant amplitude we can trivially evaluate this integral,
with the result
\be
\sigma_{p\bar{p}\to m\Psi} = 
\frac{1}{64\pi}\frac{p_{m\, cm}}{p_{p\, cm}} 
s^{-1}
\Big\{
\sum |\mathcal{M}|^2 
\Big\} .
\label{eq:cross}
\ee

\subsection{Connecting charmonium decay and production}

The partial width 
$\Gamma_{\Psi\to m p\bar p}$
and the production cross section
$\sigma_{p\bar p \to m\Psi}$ given above are simply related in the 
constant amplitude approximation, since both are given by simple
kinematic and spin factors times the same spin-summed squared 
amplitude. 
Eliminating the common squared amplitude, we find the 
following relation between the cross section and decay width:
\be
\sigma_{p\bar p\to m\Psi}=
4\pi^2 (2S_{\Psi}+1)\, 
\frac{M_{\Psi}^3}{A_D}\,
\Gamma_{\Psi\to m p\bar p}\,
\bigg[ 
\frac{p_{m\, cm}}{p_{p\, cm}} s^{-1}
\bigg].
\label{eq:connect}
\ee
Note that in this approximation all the energy dependence 
in the cross section comes from the expression in square brackets.
A similar approach has been used to estimate charmonium
cross sections of relevance to QGP experiments~\cite{Barnes:2003kp}. 

\subsection{Experimental $\Psi \to m p\bar p$ decay widths}

The estimate of the charmonium production cross sections 
$\sigma_{p\bar p\to m\Psi}$ in 
Eq.(\ref{eq:connect}) 
requires the three body decay partial widths
$\Gamma_{\Psi\to mp\bar p}$. These branching fractions
(and hence partial widths) have been measured for 
about 10 $m p\bar p$ decay modes to date, all from
$J/\psi$ or $\psi'$ initial states at $e^+e^-$ colliders. 
PDG values for the branching fractions are given in
Table~\ref{tab:widths}. Since the PDG compilation,
new measurements of $J/\psi$ and $\psi'$ branching fractions to
$mp\bar p$ have been reported by
BESII \cite{Ablikim:2005ir}
and
CLEO-c \cite{Briere:2005rc};
these are summarized in Table~\ref{tab:widthsnew}.

Unfortunately, branching fractions
of $\J^{\P\C}\neq 1^{--}$ charmonia to $mp\bar p$ final states
have not yet been reported. Results for C = $(+)$ states such as the
$\eta_c$ and $\chi_{cJ}$ would be very interesting, as these 
may couple to $mp\bar p$ through a different mechanism than 
C = $(-)$ charmonia, perhaps with significantly larger amplitudes.
This possibility is suggested by the known four-body partial widths
to $\pi^+\pi^-p\bar p$, which are larger on average for the 
$\chi_J$ states than for the $J/\psi$ and $\psi'$. 

\subsection{Previous Estimates}

The only previous estimate of near-threshold associated charmonium 
production cross sections we have found in the literature is 
the PCAC calculation of Gaillard, Maiani and Petronzio
\cite{Gaillard:1982zm}. This reference considers the process
$\Psi \to \pi^0 p \bar p$, where $\Psi$ is a generic charmonium state,
and notes that the cross section
is proportional to the peak on-resonance cross section for 
$p \bar p \to \Psi \to all$ 
if one assumes that pion emission in the decay takes place from
the $p$ or $\bar p$ line in a two-step process, 
$\Psi \to p \bar p \to \pi^0 p \bar p$. Assumption of the usual
PCAC coupling between nucleons and pions gives a relation between
the two cross sections, 
\be
\sigma_{p\bar p \to \pi^0 \Psi} =
\frac{1}{4\pi} 
\bigg(
\frac{g_A}{f_{\pi}}
\bigg)^2
\Gamma_{\Psi}^{tot}
\sigma_{CP}\,
\sigma^{max}_{p\bar p \to \Psi \to all}\, p_{\pi\, cm}.
\label{eq:PCAC_csec}
\ee
$\sigma_{CP}$ 
is a dimensionless function of the {\it c.m.} energy 
and the initial $p\bar p$ quantum numbers.
Gaillard~{\it et~al.} estimated $\sigma_{CP}$ numerically 
at a single pion energy, $E_{\pi} = 230$~MeV, 
but did not evaluate it analytically. 
We have carried out this calculation using the angular distributions 
given in their Eqs.(25-28). Defining 
$\xi_{m,n} = (1/4) \int^1_{-1} d\cos(\theta)\, \sigma_{m,n}$ 
analogous to their Eq.(38) for $\sigma_{CP}$, we find
\be
\xi_{0,0} =
1 + \frac{1}{\gamma^2} 
- \frac{\gamma_{\pi}^2 + 1}{\gamma^2}  f(\beta), 
\label{eq:xi_00}
\ee
\be
\xi_{1,0} =
-\frac{1}{2\gamma^2 (1-\beta^2)}
+\frac{1}{\beta_p^2} 
-\Big(
\frac{1}{\gamma_p^2 - 1}
+
\frac{1}{2\gamma^2}
\Big)
\, f(\beta),
\label{eq:xi_10}
\ee

\be
\xi_{1,1} =
-\frac{1}{2\gamma^2 (1-\beta^2)}
-\frac{1}{2(\gamma_p^2 -1)}
+\frac{1+\beta_p^2}{2(\gamma_p^2 -1)} 
\, f(\beta)
\label{eq:xi_11}
\ee
where 
$\beta = \beta_{\pi}  \beta_p$,
$\gamma = \gamma_{\pi}  \gamma_p$
and 
$f(x) = \tanh^{-1}(x)/x$.
The coupling $\sigma_{CP}$ is a linear combination of these 
$\xi$ functions. For example, for a pure S-wave $\pi^0 J/\psi$ system
one requires a CP$ = -$, $^1$P$_1$ $1^{+-}$ $p\bar p$ initial state, 
for which $\sigma_{CP = -} = \xi_{0,0}$.
For a CP$ = +$, spin-triplet initial state, $\sigma_{CP = +}$ is a linear 
combination of $\xi_{1,0}$ and $\xi_{1,1}$,
\be
\sigma_{CP = +} = 
(1 - \langle S_3^2 \rangle ) \, \xi_{1,0} +
\langle S_3^2 \rangle \, \xi_{1,1}. 
\label{eq:PCAC_sigma}
\ee 
The spin matrix element $\langle S_3^2 \rangle$
depends on the mix of ${\L} = {\J} \pm 1$ waves in the initial 
$p\bar p$ state, and satisfies $0 \leq \langle S_3^2 \rangle \leq 1$. 
(For an initial ${}^3$S$_1$ $p\bar p$ state, 
$\langle S_3^2 \rangle = 2/3$.) 

Numerical results for $\sigma_{p\bar p \to \pi^0 J/\psi}$
using both this PCAC approach and the constant amplitude formula
of Eq.(\ref{eq:connect}) are given below.

\section{Numerical results for $\sigma_{p\bar p\to m \Psi}$}

\subsection{Method}

Using Eq.(\ref{eq:connect}) and the partial widths given in
Tables~\ref{tab:widths} and \ref{tab:widthsnew}, we have generated
constant amplitude estimates 
for $\sigma_{p\bar p \to m \Psi}$ for all cases with measured 
$\Psi$ partial widths to $mp\bar p$ final states. 
The $J/\psi$ cross section estimates use PDG branching fractions and 
decay widths, and the $\psi'$ cross sections were calculated using 
BESII branching fractions for $m=\pi^0$ and $\eta$ and 
CLEO-c branching fractions for $\rho$, $\omega$ and $\phi$.
Our results, including combined errors from the total widths and 
branching fractions,
are given in Table~\ref{tab:calculations}.

\subsection{The case $p\bar p\to \pi^0 J/\psi$}

The process $p\bar{p}\to J/\psi \pi^0$ is of special interest
because this is the single associated charmonium production 
reaction in $p\bar p$ annihilation for which we have 
an experimental measurement of the cross section.
This reaction was studied by the E760 Collaboration at Fermilab 
\cite{Armstrong:1992ae}, who observed a few $p\bar p\to \pi^0 J/\psi$ 
events as a background in their searches for 
${}^1$P$_1$ $h_c$ and $2{}^3$S$_1$ $\eta_c'$ charmonia.

\begin{figure}[ht]
\vskip 1.0cm
\begin{center}
\includegraphics[width=0.9\linewidth]{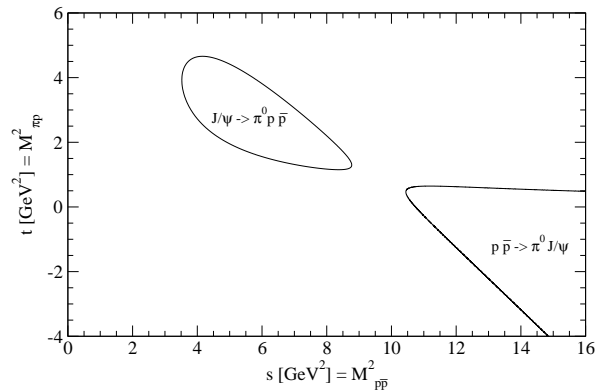}
\caption{Kinematically allowed regions for the three-body decay
$J/\psi \to \pi^0 p\bar p$ and the related
charmonium production reaction
$p\bar p \to \pi^0 J/\psi$.}
\label{fig:Dalitz1}
\end{center}
\end{figure}

\begin{figure}[ht]
\begin{center}
\includegraphics[width=0.9\linewidth]{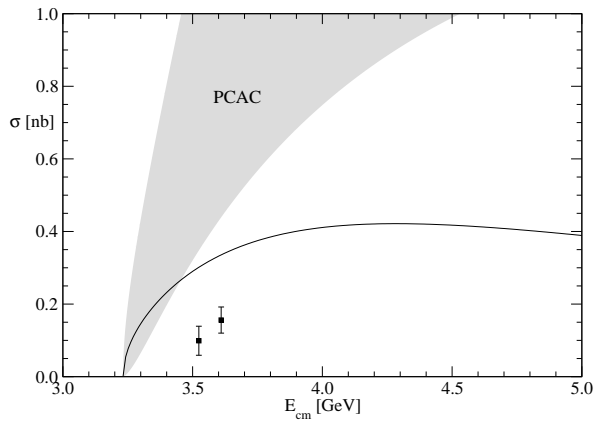}
\caption{Theoretical and experimental 
cross sections for $p\bar p\to \pi^0 J/\psi$. 
The theoretical predictions are the constant amplitude result 
Eq.(\ref{eq:connect}) (solid) and the range of PCAC cross sections,
from Eq.(\ref{eq:PCAC_csec}) (filled).
The experimental points are from E760 \cite{Armstrong:1992ae}.}
\label{fig:csec}
\end{center}
\end{figure}

The observed events correspond to a cross
section of $99\pm 40\, pb$ just below 3.525~GeV and 
$156\pm 36\, pb$ at 3.61~GeV. 
This has also been quoted as a combined value of
$130\pm 25\, pb$ \cite{Cester}. 
The E760 measurements are shown in
Fig.\ref{fig:csec} together with the theoretical predictions.

The constant amplitude cross section 
of Eq.(\ref{eq:connect}) is also shown in Fig.\ref{fig:csec}. 
This approximation 
predicts $\sigma = 299\, pb$ at 3.52~GeV and $336\, pb$ at 3.61~GeV, which
overestimates the cross section at these energies 
by about a factor of 2-3.
This suggests that the 
constant amplitude approximation is useful as a first estimate, 
but should be improved through a more detailed description of the
amplitude, for example by incorporating the contributions of 
individual resonances to the decay Dalitz plot. 
The importance of baryon resonances in this decay 
and in related
$\Psi \to mp\bar p$ decays has been discussed previously in
both 
experimental \cite{Bai:2001ua,Ablikim:2005ir,Ablikim:2004ug}
and 
theoretical \cite{Sinha:1984qn,Liang:2004sd}
references.

The PCAC result for the cross section for
$J/\psi \to \pi^0  p \bar p$, 
Eq.(\ref{eq:PCAC_csec}),
is also shown in Fig.\ref{fig:csec}. 
Here we have assumed (rounded) PDG values of 
$g_A = 1.27$ and $f_{\pi} = 0.131$~GeV,
PDG masses and widths, and a 
peak on-resonance cross section of
$\sigma^{max}_{p\bar p \to J/\psi \to all} = 5.0\, \mu b$ \cite{csec_max}.
Since the result depends on the initial $p\bar p$ quantum numbers assumed,
we show the full set of values that follow from the range
$0 \leq \langle S^2_3 \rangle \leq 1$ in 
Eq.(\ref{eq:PCAC_sigma}).
Evidently this PCAC prediction is also somewhat larger than the experimental 
E760 cross section in the 3.5-3.6 GeV region where we have data, and is 
rather similar to the constant amplitude result 
near threshold. 
The disagreement of PCAC with experiment may simply imply 
that other important processes interfere destructively 
with the nucleon pole diagram. 

\subsection{Other channels}

Our constant amplitude predictions for all 
$p\bar p \to m\Psi$ cross sections are summarized in
Table~\ref{tab:calculations}. 

\begin{figure}[ht]
\vskip 0.8cm
\begin{center}
\includegraphics[width=0.9\linewidth]{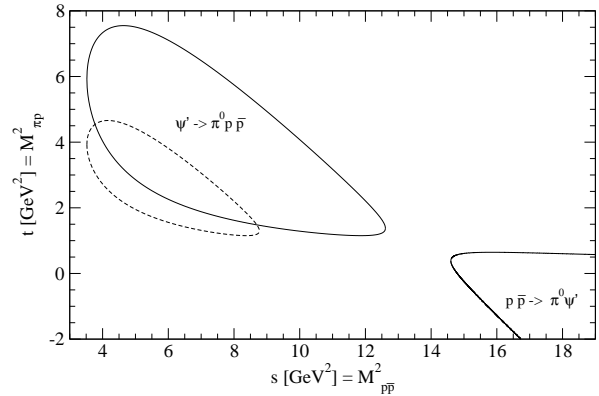}
\caption{Allowed regions for $\psi' \to \pi^0 p\bar p$ and
$p\bar p \to \pi^0 \psi' $. The smaller physical region for
$J/\psi \to \pi^0 p\bar p $ is shown for comparison (dashed, 
from Fig.\ref{fig:Dalitz1}).}
\label{fig:Dalitz2}
\end{center}
\end{figure}

The $p \bar p \to m\psi'$ cross sections are typically predicted to be
10s of $nb$, at least an order of magnitude smaller than the corresponding
$J/\psi$ cross sections. There is no indication of an enhancement of
isoscalar $m$ final states in this case. Although the $\psi'$ couplings are
relatively weak, the decay $\psi' \to \pi^0 p \bar p$ (and the 
corresponding reaction $p\bar p \to \pi^0 \psi'$) 
may be the most interesting to study in the near future. 
As is evident in Fig.\ref{fig:Dalitz2}, the region of
phase space sampled by this decay is much larger than in
$J/\psi \to \pi^0 p \bar p$, so more baryon 
resonance contributions can be investigated, and their effects
on the extrapolation to the physical region for
$p\bar p \to \pi^0 \psi'$ 
can be studied once data on this reaction becomes available.
BESII \cite{Ablikim:2005ir} has recently reported 
results on this $\psi'$ decay mode.

\begin{figure}[ht]
\vskip 1.0cm
\begin{center}
\includegraphics[width=0.9\linewidth]{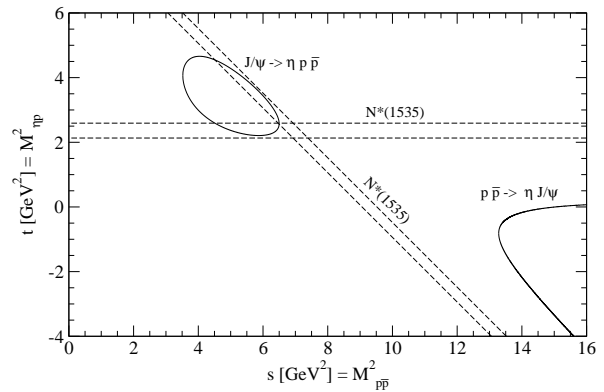}
\caption{Allowed regions for $J/\psi \to \eta p\bar p$ and
$p\bar p\to \eta J/\psi$, showing N$^*(1535)$ bands.
(M$\pm \Gamma/2$ limits are shown.)}
\label{fig:Dalitz3}
\end{center}
\end{figure}

\eject
It is notable that 
quite large maximum cross sections of $1$-$3~nb$ 
are predicted for 
$\Psi = J/\psi$ and $m = \eta, \omega, \eta'$. 
This suggests that it may be interesting for PANDA to
consider reactions in which $m$ is a light isoscalar
meson. This is also supported by the relatively large
branching fractions observed for  
charmonium decays to $\pi^+ \pi^- p \bar p $.
Although an enhancement of 
$p\bar p \to m \Psi$
cross sections
with light isoscalar $m$ 
is an interesting possibility, one should note that the 
Dalitz plot for the decay $J/\psi \to  \eta p \bar p$ (as one example)
is crossed by N$^*(1535)$ bands (Fig.\ref{fig:Dalitz3}), 
and this baryon resonance contribution
may be the reason for the large branching fraction. 
This is supported by results from BESII \cite{Bai:2001ua}.
As these N$^*(1535)$ bands lie far outside the physical region for 
$p\bar p \to \eta J/\psi$, it appears likely that 
the constant amplitude approximation Eq.(\ref{eq:connect})
will seriously overestimate this cross section. 

\section{Summary and Future}

We have used the partial decay widths of charmonia into $mp\bar p$
final states (where $m$ is a light meson) and a constant amplitude
approximation to estimate cross sections for the reactions 
$p\bar p\to m\Psi$, which are of great interest for the planned 
PANDA experiment. The single experimentally measured cross section
of this type is $p\bar p\to \pi^0 J/\psi$, which was reported by E760
to be $130\pm 25~pb$ near 3.5-3.6 GeV. In comparison we predict 
a cross section of about $300$~$pb$ at this energy. 
Application of this constant amplitude approximation to other 
$p\bar p \to m J/\psi$ reactions suggests enhanced peak cross sections 
of as large as 1-3~$nb$ for light isoscalar $m$. This prediction 
may however be an artifact of large baryon 
resonance contributions to these specific $J/\psi \to mp\bar p$ 
decays. These resonances need not contribute significantly 
to the production cross section if the decay resonance bands lie far 
from that region of phase space
(see for example Fig.\ref{fig:Dalitz3}). 
The corresponding $\psi'$ decays do not show evidence of 
an enhancement of isosinglet $m$ final states; the estimated 
$p\bar p\to m\psi'$ cross sections are 10s of $nb$.
 
A next important step in these calculations will be to
model the contributions of individual baryon resonances to the decays.
In low-energy processes involving a single $\pi^0$, such as 
$J/\psi \to  \pi^0 p \bar p$ 
and 
$p \bar p \to \pi^0 J/\psi$
(Fig.\ref{fig:Dalitz1}), one may find dominance of the decay
Dalitz plot by a few baryon resonances. 
In such cases, extrapolation of the amplitude to the reaction region 
may be relatively straightforward. 
Studies of the corresponding $\psi'$ processes may be instructive, 
since the region of phase space accessible to the decay 
is much larger (Fig.\ref{fig:Dalitz2}), 
and new data is expected from BESII and CLEO-c. 
In contrast, relating decays and production cross sections for
heavier mesons such as the $\eta$ is clearly problematic, since the 
decay Dalitz plot provides a smaller window on the amplitude, and 
the extrapolation from decay to production covers a much larger kinematic 
range (Fig.\ref{fig:Dalitz3}). For vector mesons such as the $\omega$ 
we encounter the additional complication that the couplings of 
baryon resonances to $\omega$N are not well established. 
The N$^*$ experimental program at TJNAF, which has a goal of 
determining these $\omega$N and related baryon resonance decay
couplings, may also be useful in interpreting $\Psi \to mp\bar p$ 
decays and clarifying their relation to charmonium production 
cross sections. 

Future measurements of $mp\bar p$ branching fractions
of C~=~(+) charmonia such as the $\{\chi_J\}$ and $\eta_c$ 
will be especially interesting, as these will allow 
estimates of their $p\bar p \to m\Psi$ production cross sections. 
As C~=~(+) charmonia employ different production 
mechanisms than the C = ($-$) states, for example $gg$
rather than $ggg$ intermediaries, they may have quite 
different production cross sections. 

\section{Acknowledgements}

We are happy to acknowledge useful communications with
D.Bettoni and C.Patrignani regarding E760 and E835 measurements
of charmonium production, R.Galik, G.S.Huang, H.Mahlke-Kr\"uger,
T.Pedlar and J.Yelton regarding CLEO studies of charmonium decays to 
$mp\bar p$ final states, G.F\"aldt and C.Wilkin regarding soft pion 
processes, and M.K.Gaillard regarding the PCAC cross section calculation. 
This research was supported in part by the Centre for 
Dynamical Processes and Structure Formation at the 
Faculty of Science and Technology and the Physics Graduate School 
gradU at Uppsala University, the Swedish Science Council VR, 
the U.S. National Science Foundation through grant NSF-PHY-0244786 at the
University of Tennessee, and the U.S. Department of Energy under contract
DE-AC05-00OR22725 at Oak Ridge National Laboratory.

\vskip 1cm

\begin{table}[ht]
\begin{tabular}{|l|c|r|}\hline
\qquad Decay 
& \hskip5mm $\Gamma_{mp\bar p}$ [eV] \hskip5mm
& Experiment $\hskip5mm$ \\ \hline
$\,$
$J/\psi\to\pi^0 p\bar p$  & $99.2 \pm 8.9$ & MRK2, MRK1, DASP \\
$\,$
$J/\psi\to\eta\, p\bar p$   & $190\pm 18$    & MRK2, MRK1, DASP \\
$\,$
$J/\psi\to\rho^0 p\bar p$   & $< 28$        & MRK2 \\
$\,$
$J/\psi\to\omega\, p\bar p$ & $118\pm 23$    & MRK2, MRK1 \\
$\,$
$J/\psi\to\eta' p\bar p$  & $82\pm 37$     & MRK2, MRK1 \\
$\,$
$J/\psi\to\phi\, p\bar p$   & $4.1\pm 1.4$   & DM2 \\
\hline
\quad $\psi'\ \to\pi^0 p\bar p$  & $39\pm 14$     & MRK2 \\
\quad $\psi'\ \to\omega\, p\bar p$ & $22 \pm 9$     & BES \\
\quad $\psi'\ \to\phi\, p\bar p$   & $ < 7.4$         & BES \\ 
\hline
\end{tabular}
\caption{Experimental charmonium partial widths 
to $m p\bar p$ final states. 
These numbers combine the PDG \cite{PDG} total widths with
the reported branching fractions.}
\label{tab:widths}
\end{table}

\vfill\eject

\begin{table}
\begin{tabular}{|l|c|c|}
\hline
\quad Decay 
&  $\Gamma_{mp\bar p}^{\rm (BES)}$[eV] 
&  $\Gamma_{mp\bar p}^{\rm (CLEO)}$[eV]   \\ 
\hline
\
$\psi'\to\pi^0 p\bar p$  & \ $37.1 \pm 5.5$ \  & -      \\
\
$\psi'\to \eta \, p\bar p$   & $16.3\pm 3.8$   & $22\pm 12$ \\
\ 
$\psi'\to\rho^0 p\bar p$   & -               & $14\pm 6$  \\
\
$\psi'\to\omega\, p\bar p$ & -               & $17\pm 8$ \\
\
$\psi'\to\phi\, p\bar p$   & -               & $ < 6.7$   \\ 
\hline
\end{tabular}
\caption{$\psi' \to mp\bar p$ partial widths recently
reported by BESII \cite{Ablikim:2005ir} 
and CLEO-c \cite{Briere:2005rc}.
These numbers combine the PDG \cite{PDG} total width with
the reported branching fractions.}
\label{tab:widthsnew}
\end{table}

\begin{table}[ht]
\begin{center}
\begin{tabular}{|l|c|c|c|}
\hline
\quad Reaction
& $\sigma^{max}_{p\bar p\to m\Psi}$[$pb$] 
& E$_{cm}^{max}$[GeV]
& $A_D$[GeV$^4$] 
\\ 
\hline
\
$p\bar p \to \pi^0 J/\psi $ 
& $420\pm 40$               
& 4.28 
& \phantom{1}9.265 
\\
\
$p\bar p \to \eta\, J/\psi  $   
& $1520\pm 140$               
& 4.57  
& \phantom{1}4.520 
\\
\
$p\bar p \to\rho^0 J/\psi $   
& $ < 450$               
& 4.80  
& \phantom{1}2.114 
\\
\
$p\bar p \to\omega\, J/\psi $ 
& $1900\pm 400$               
& 4.80  
& \phantom{1}2.053 
\\
\
$p\bar p \to\eta' J/\psi $  
& $3300\pm 1500$              
& 4.99  
& \phantom{1}0.765 
\\
\
$p\bar p \to\phi\, J/\psi  $   
& $280\pm 90$                
& 5.06  
& \phantom{1}0.452 
\\
\hline
\
$p\bar p \to\pi^0 \psi' $    
& $55\pm 8$     
& 5.14  
& 30.500
\\
\
$p\bar p \to\eta\, \psi'$     
& $33\pm 8$     
& 5.38  
& 20.984
\\
\
$p\bar p \to\rho^0 \psi' $   
& $38\pm 17$    
& 5.59  
& 14.953
\\
\
$p\bar p \to\omega\, \psi' $   
& $46\pm 22$   
& 5.60  
& 14.778
\\
\
$p\bar p \to\phi\, \psi' $     
& $ < 28$       
& 5.84  
& \phantom{1}9.118
\\
\hline
\end{tabular}
\caption{Charmonium cross sections from 
Eq.(\ref{eq:connect}),
given the $\Psi \to mp\bar p $ partial widths 
of Tables~\ref{tab:widths}~and~\ref{tab:widthsnew}.
The size and energy of the cross section maximum and the area
of the $\Psi \to mp\bar p $ Dalitz plot are given. 
These cross sections may be overestimated 
if the corresponding $\Psi$ decay Dalitz plot is dominated 
by baryon resonances (see text).}
\label{tab:calculations}
\end{center}
\end{table}

\end{document}